\documentstyle[12pt]{article}
\newcommand{\bfp}{\mbox{\boldmath $p$}}
\newcommand{\bfP}{\mbox{\boldmath $P$}}
\newcommand{\bfk}{\mbox{\boldmath $k$}}
\def\nostrocostruttino#1\over#2{\mathrel{\mathop{\kern 0pt \rlap
{\hbox{$#1$}}} \hbox{\kern-.135em $#2$}}}
\def\sumint{\nostrocostruttino \sum \over {\displaystyle\int}}
\newcommand{\NP}[1]{{\it Nucl.\ Phys.}\ {\bf #1}}
\newcommand{\ZP}[1]{{\it Z.\ Phys.}\ {\bf #1}}
\newcommand{\PL}[1]{{\it Phys.\ Lett.}\ {\bf #1}}
\newcommand{\PR}[1]{{\it Phys.\ Rev.}\ {\bf #1}}

\def\lsim{\mathrel{\rlap{\lower4pt\hbox{\hskip1pt$\sim$}}\raise1pt\hbox{$<$}}}
\def\gsim{\mathrel{\rlap{\lower4pt\hbox{\hskip1pt$\sim$}}\raise1pt\hbox{$>$}}}
\begin{document}
\begin{flushright}
DFTT 61/95 \\
INFNCA-TH9526 \\
hep-ph/9512379 \\
\end{flushright}
\vskip 1.5cm
\begin{center}
{\bf
Polarized inclusive leptoproduction, $\ell N \to hX$, and the
hadron helicity density matrix, $\rho(h)$: possible measurements
and predictions
}\\
\vskip 1.5cm
{\sf M.\ Anselmino$^1$, M.\ Boglione$^1$, J.\ Hansson$^2$ and F.\ Murgia$^3$}
\vskip 0.8cm
{$^1$Dipartimento di Fisica Teorica, Universit\`a di Torino and \\
      INFN, Sezione di Torino, Via P. Giuria 1, 10125 Torino, Italy\\
\vskip 0.5cm
 $^2$Department of Physics, Lule{\aa} University of Technology, S-97187
      Lule\aa, Sweden \\
\vskip 0.5cm
 $^3$INFN, Sezione di Cagliari, Via A. Negri 18, 09127 Cagliari, Italy } \\
\end{center}
\vskip 1.5cm
\noindent
{\bf Abstract:} \\
We discuss the production of hadrons in polarized lepton nucleon
interactions and in the current jet fragmentation region; using the
QCD hard scattering formalism we compute the helicity density matrix
of the hadron and show how its elements, when measurable, can give
information on the spin structure of the nucleon and the spin dependence
of the quark fragmentation process. The cases of $\rho$ vector mesons
and $\Lambda$ baryons are considered in more details and, within
simplifying assumptions, some estimates are given.
\newpage
\pagestyle{plain}
\setcounter{page}{1}
\noindent
{\bf 1 - Introduction and general formalism}
\vskip 6pt
The full description of hard scattering processes involving hadrons
always requires a knowledge of both the elementary interactions between
the hadronic constituents and the constituent distribution or fragmentation
properties; while the former interactions are computable in perturbative
QCD or QED the latter properties, {\it i.e.} the amount of quarks and gluons
inside hadrons and the amount of observed particles resulting from a quark
or gluon fragmentation, are non perturbative and cannot be computed in QCD.
However, their universality and the QCD knowledge of their $Q^2$ evolution
allow, once some information is obtained from certain processes, to use it
in other processes in order to make genuine predictions. It is then crucial
to collect phenomenological information on these non perturbative quantities.

On the quark and gluon content of the nucleons we have by now gathered
plenty of detailed information mainly from unpolarized Deep Inelastic
lepton-nucleon Scattering; some information is also available on
unpolarized quark fragmentation properties either from DIS or $e^-e^+$
annihilations. Much less we know about the inner structures of polarized
hadrons and their dynamical properties. The proton and neutron spin structure
functions have recently received much attention and their improved
measurements have caused great surprise and enormous theoretical activity
\cite{ael}, but we still need a better knowledge; very little is known on
polarized quark and gluon fragmentations. Several observed spin effects are
not well understood and are certainly related to non perturbative hadronic
properties.

We consider here the inclusive deep inelastic process
\begin{equation}
\ell N \to h X
\label{pro}
\end{equation}
in which an unpolarized or polarized lepton scatters off a polarized nucleon
and one observes a final hadron $h$ whose spin state is studied through the
measurement of its helicity density matrix $\rho(h)$. The incoming lepton
interacts with a polarized quark inside the polarized nucleon and the
quark then fragments into the hadron $h$ contributing to its spin;
thus, we expect to learn something on the polarized quark distribution
and fragmentation functions. We consider spin 1 and spin 1/2 final hadrons
and different polarizations of the initial nucleon; we consider either
unpolarized or longitudinally polarized leptons because, as we shall see,
transversely polarized ones cannot add any further information.

According to the QCD hard scattering scheme and the factorization theorem
\cite{col1}-\cite{col3}, \cite{old} the helicity density matrix of the hadron
$h$ inclusively produced in reaction (\ref{pro}) is given by
\begin{eqnarray}
\rho_{\lambda^{\,}_h,\lambda^\prime_h}^{(s,S)}(h) \>
\frac{E_h \, d^3\sigma^{\ell,s + N,S \to h + X}} {d^{3} \bfp_h} &=&
\sum_{q; \lambda^{\,}_{\ell}, \lambda^{\,}_q, \lambda^\prime_q}
\int \frac {dx}{\pi z} \frac {1}{16 \pi x^2 s^2} \label{gen} \\
& & \!\!\!\! \rho^{\ell,s}_{\lambda^{\,}_{\ell}, \lambda^{\,}_{\ell}} \,
\rho_{\lambda^{\,}_q, \lambda^{\prime}_q}^{q/N,S} \, f_{q/N}(x) \,
\hat M^q_{\lambda^{\,}_{\ell}, \lambda^{\,}_q;
\lambda^{\,}_{\ell}, \lambda^{\,}_q} \,
\hat M^{q*}_{\lambda^{\,}_{\ell}, \lambda^{\prime}_q;
\lambda^{\,}_{\ell}, \lambda^{\prime}_q} \,
D_{\lambda^{\,}_h, \lambda^{\prime}_h}^{\lambda^{\,}_q,\lambda^{\prime}_q}(z)
\nonumber
\end{eqnarray}
where $\rho^{\ell,s}$ is the helicity density matrix of the initial
lepton with spin $s$, $f_{q/N}(x)$ is the number density of unpolarized
quarks $q$ with momentum fraction $x$ inside an unpolarized nucleon and
$\rho^{q/N,S}$ is the helicity density matrix of quark $q$ inside the
polarized nucleon $N$ with spin $S$.
The $\hat M^q_{\lambda^{\,}_{\ell}, \lambda^{\,}_q; \lambda^{\,}_{\ell},
\lambda^{\,}_q}$'s are the helicity amplitudes for the elementary process
$\ell q \to \ell q$. The final lepton spin is not observed and helicity
conservation of perturbative QCD and QED has already been taken into account
in the above equation: as a consequence only the diagonal elements of
$\rho^{\ell,s}$ contribute to $\rho(h)$ and non diagonal elements, present in
case of transversely polarized leptons, do not contribute.
$D^{\lambda^{\,}_q,\lambda^{\prime}_q}_{\lambda^{\,}_h,\lambda^{\prime}_h}(z)$
is the product of {\it fragmentation amplitudes}
\begin{equation}
D^{\lambda^{\,}_q,\lambda^{\prime}_q}_{\lambda^{\,}_h,\lambda^{\prime}_h}(z)
=\sumint_{X,\lambda^{\,}_X} {\cal D}_{\lambda^{\,}_{X}, \lambda_h;
\lambda^{\,}_q} \,
{\cal D}^*_{\lambda^{\,}_{X}, \lambda^{\prime}_h; \lambda^{\prime}_q}
\label{framp}
\end{equation}
where the $\sumint_{X,\lambda^{\,}_X}$ stands for a spin sum and phase space
integration of the undetected particles, considered as a system $X$.
The usual unpolarized fragmentation function $D_{h/q}(z)$, {\it i.e.}
the density number of hadrons $h$ resulting from the fragmentation of
an unpolarized quark $q$ and carrying a fraction $z$ of its momentum,
is given by
\begin{equation}
D_{h/q}(z) = {1\over 2} \sum_{\lambda^{\,}_q, \lambda^{\,}_h}
D^{\lambda^{\,}_q,\lambda^{\,}_q}_{\lambda^{\,}_h,\lambda^{\,}_h}(z)
= {1\over 2} \sum_{\lambda^{\,}_q, \lambda^{\,}_h}
D_{h_{\lambda^{\,}_h}/q_{\lambda^{\,}_q}}(z) \,,
\label{fr}
\end{equation}
where $D^{\lambda^{\,}_q,\lambda^{\,}_q}_{\lambda^{\,}_h,
\lambda^{\,}_h}(z) \equiv D_{h_{\lambda^{\,}_h}/q_{\lambda^{\,}_q}}$ is
a polarized fragmentation function, {\it i.e.} the density number of
hadrons $h$ with helicity $\lambda^{\,}_h$ resulting from the fragmentation
of a quark $q$  with helicity $\lambda^{\,}_q$. Notice that
by definition and parity invariance the generalized fragmentation
functions (\ref{framp}) obey the relationships
\begin{eqnarray}
D^{\lambda^{\,}_q,\lambda^{\prime}_q}_{\lambda^{\,}_h,\lambda^{\prime}_h}
&=& \left( D^{\lambda^{\prime}_q,\lambda^{\,}_q}
_{\lambda^{\prime}_h,\lambda^{\,}_h} \right)^* \, \label{def} \\
D^{-\lambda^{\,}_q,-\lambda^{\prime}_q}_{-\lambda^{\,}_h,-\lambda^{\prime}_h}
&=& -(-1)^{2S_h}(-1)^{\lambda^{\,}_q + \lambda^{\prime}_q
+ \lambda^{\,}_h + \lambda^{\prime}_h} \>
D^{\lambda^{\,}_q,\lambda^{\prime}_q}_{\lambda^{\,}_h,\lambda^{\prime}_h} \,,
\label{par}
\end{eqnarray}
where $S_h$ is the hadron spin; notice also that collinear configuration
(intrinsic $\bfk_{\perp} =0$) together with angular
momentum conservation in the forward fragmentation process imply
\begin{equation}
D^{\lambda^{\,}_q,\lambda^{\prime}_q}_{\lambda^{\,}_h,\lambda^{\prime}_h}
= 0 \quad\quad {\mbox{\rm when}} \quad\quad
\lambda^{\,}_q - \lambda^{\prime}_q \not= \lambda^{\,}_h-\lambda^{\prime}_h \,.
\label{for}
\end{equation}

Eq. (\ref{gen}) holds at leading twist, leading order in the
coupling constants and large $Q^2$ values; the intrinsic $\bfk_\perp$
of the partons have been integrated over and collinear configurations
dominate both the distribution functions and the fragmentation processes.
For simplicity of notations we have not indicated the $Q^2$ scale
dependences in $f$ and $D$; the variable $z$ is related to
$x$ by the usual imposition of energy momentum conservation
in the elementary 2 $\to$ 2 process \cite{bla}:
\begin{equation}
z= -{t+xu \over xs}
\label{zet}
\end{equation}
where $s,t$ and $u$ are the Mandelstam variables for the $\ell N \to hX$
process, related to the Mandelstam variables $\hat s, \hat t$ and $\hat u$
for the subprocess $\ell q \to \ell q$ by
\begin{equation}
\hat s = xs \quad\quad  \hat t = xu/z = -Q^2 \quad\quad \hat u = t/z \,.
\label{man}
\end{equation}
Finally, the elementary amplitudes $\hat M^q$ are normalized so that
\begin{equation}
{d\hat\sigma^q \over d\hat t} = {1\over 16\pi \hat s^2} \>
{1 \over 4} \, \sum_{\lambda^{\,}_{\ell}, \lambda^{\,}_q}
|\hat M^q_{\lambda^{\,}_{\ell}, \lambda^{\,}_q;
\lambda^{\,}_{\ell}, \lambda^{\,}_q}|^2
\label{norm}
\end{equation}
is the unpolarized elementary cross-section and the normalization factor
in the left hand side of Eq. (\ref{gen}), such that Tr$\rho(h)=1$,
is the cross-section for the inclusive production of a hadron $h$, summed
over all possible hadron helicities, in the DIS scattering of a lepton with
spin $s$ off a nucleon with spin $S$:
\begin{equation}
\frac{E_h \, d^3\sigma^{\ell,s + N,S \to h + X}} {d^{3} \bfp_h} =
\sum_{q; \lambda^{\,}_{\ell}, \lambda^{\,}_q}
\int \frac {dx}{16 \pi^2 x^2 z s^2}\>
\rho^{\ell,s}_{\lambda^{\,}_{\ell}, \lambda^{\,}_{\ell}} \,
\rho_{\lambda^{\,}_q, \lambda^{\,}_q}^{q/N,S} \, f_{q/N}(x) \,
|\hat M^q_{\lambda^{\,}_{\ell}, \lambda^{\,}_q;
\lambda^{\,}_{\ell}, \lambda^{\,}_q}|^2 D_{h/q}(z)
\label{d3s}
\end{equation}
where we have made use of Eqs. (\ref{fr}), (\ref{par}) and (\ref{for}).
In the sequel we shall adopt the short notation:
\begin{equation}
{\hat M^q_{++;++} \over 4 \sqrt {\pi} \, \hat s} \equiv \hat M^q_+ \quad\quad
{\hat M^q_{+-;+-} \over 4 \sqrt {\pi} \, \hat s} \equiv \hat M^q_- \,.
\label{not}
\end{equation}

A measurement of $\rho(h)$ allows, via Eq. (\ref{gen}) and the knowledge
of the unpolarized distribution functions $f_{q/N}(x)$ and of the elementary
lepton-quark interaction, to obtain new information on the spin dependent
quark distribution and fragmentation processes. The quark helicity density
matrix $\rho^{q/N,S}$ can be decomposed as
\begin{equation}
\rho^{q/N,S} = P_P^{q/N,S} \rho^{N,S} + P_A^{q/N,S} \rho^{N,-S}
\label{rhoq}
\end{equation}
where $P_{P(A)}^{q/N,S}$ (which, in general, depends on $x$)
is the probability that the spin of the quark
inside the polarized nucleon $N$ is parallel (antiparallel) to the nucleon
spin $S$ and $\rho^{N,S(-S)}$ is the helicity density matrix of the nucleon
with spin $S(-S)$. Notice that
\begin{equation}
P^{q/N,S} = P_P^{q/N,S} - P_A^{q/N,S}
\label{polq}
\end{equation}
is the component of the quark polarization vector along the parent nucleon
spin direction.

In the next Sections we shall consider several particular cases of
Eq. (\ref{gen}) and discuss what can be learned or expected from a
measurement of $\rho(h)$. For clarity and completeness of the discussion
we also report or rederive known results. Towards completion of this work
a most general analysis of polarized DIS leptoproduction has appeared in the
literature \cite{mul}, taking into account ${\cal O}(1/Q)$ corrections;
our work puts more emphasis on possible measurements, gives numerical
estimates and deals also with spin 1 final hadrons.

\goodbreak
\vskip 12pt
\noindent
{\bf 2 - Spin 1 final hadron; unpolarized leptons and polarized nucleons}
\vskip 6pt
\nobreak

We consider first the production of a spin 1 hadron ($h=V$) with unpolarized
leptons,
\begin{equation}
\rho_{\lambda^{\,}_{\ell},\lambda^{\prime}_{\ell}}^{\ell,s} = {1\over 2} \,
\delta_{\lambda^{\,}_{\ell},\lambda^{\prime}_{\ell}} \,,
\label{runp}
\end{equation}
and, for the moment, a generic nucleon spin $S$. Eqs. (\ref{gen}), (\ref{for})
and (\ref{norm}), together with parity invariance of
the elementary QED process, then give
\begin{eqnarray}
\rho_{1,1}^{(S)}(V) \, d^3\sigma &=&
\sum_q \int \frac {dx}{\pi z} \, f_{q/N} \, d\hat\sigma^q
\left[ \rho_{+,+}^{q/N,S} D_{1,1}^{+,+}
+ \rho_{-,-}^{q/N,S} D_{1,1}^{-,-} \right] \label{v11} \\
\rho_{0,0}^{(S)}(V) \, d^3\sigma &=&
\sum_q \int \frac {dx}{\pi z} \, f_{q/N} \, d\hat\sigma^q
\, D_{0,0}^{+,+} \label{v00} \\
\rho_{-1,-1}^{(S)}(V) \, d^3\sigma &=&
\sum_q \int \frac {dx}{\pi z} \, f_{q/N} \, d\hat\sigma^q
\left[ \rho_{+,+}^{q/N,S} D_{-1,-1}^{+,+}
+ \rho_{-,-}^{q/N,S} D_{-1,-1}^{-,-} \right] \label{v-1-1} \\
\rho_{1,0}^{(S)}(V) \, d^3\sigma &=&
\sum_q \int \frac {dx}{\pi z} \, f_{q/N}
\left[ \mbox{\rm Re} \hat M_+^q \hat M_-^{q*} \right]
\rho_{+,-}^{q/N,S} D_{1,0}^{+,-} \label{v10} \\
\rho_{-1,0}^{(S)}(V) \, d^3\sigma &=&
\sum_q \int \frac {dx}{\pi z} \, f_{q/N}
\left[ \mbox{\rm Re} \hat M_+^q \hat M_-^{q*} \right]
\rho_{-,+}^{q/N,S} D_{-1,0}^{-,+} \label{v-10}
\end{eqnarray}
where the $\pm$ indices stand respectively for $\pm 1/2$ helicities;
$d\hat\sigma^q$ stands for $d\hat\sigma^q/d\hat t$ , Eq. (\ref{norm}),
and $d^3\sigma$ is a short notation for the cross-section of Eq. (\ref{d3s})
with unpolarized leptons; it just equals the unpolarized cross-section
\begin{equation}
\frac{E_h \, d^3\sigma^{\ell N \to h X}} {d^{3} \bfp_h} =
\sum_q \int \frac {dx}{\pi z} \>
f_{q/N}(x) \, {d\hat\sigma^q \over d\hat t} \, D_{h/q}(z) \,,
\label{sunp}
\end{equation}
as can be seen from Eq. (\ref{d3s}) upon using Eqs. (\ref{norm}), (\ref{runp})
and the parity relation $\sum_{\lambda^{\,}_{\ell}}
|\hat M^q_{\lambda^{\,}_{\ell}, \lambda^{\,}_q;
\lambda^{\,}_{\ell}, \lambda^{\,}_q}|^2 = 32 \pi \hat s^2
d\hat\sigma^q/d\hat t$. By exploiting Eqs. (\ref{def}) and (\ref{par})
one can check that $\rho_{1,1} + \rho_{0,0} + \rho_{-1,-1} = 1$ and
$\rho_{0,\pm 1} = (\rho_{\pm 1,0})^*$; notice
also that, due to Eq. (\ref{for}), $\rho_{1,-1} = \rho_{-1,1} = 0$.

Eqs. (\ref{v11})-(\ref{v-10}) hold for any polarization $S$ of the nucleon
and in any reference frame; we shall now consider particular nucleon spin
configurations in the lepton-nucleon centre of mass frame. Also the numerical
results given in the last Section will be obtained in the c.m. frame.
We choose $xz$ as the hadron production plane with the lepton moving along
the $z$-axis and the nucleon in the opposite direction; as usual we indicate
by an index $L$ the (longitudinal) nucleon spin orientation along the
$z$-axis, by an index $S$ the (sideway) orientation along the $x$-axis and
by an index $N$ the (normal) orientation along the $y$-axis.

\vskip 6pt
\noindent
{\it a) Nucleon longitudinal polarization}, $S = S_L$

In this case the helicity density matrix of the nucleon is given by
(notice that $S= \pm S_L$ means, for the nucleon moving opposite to the
$z$-direction, $\lambda_N = \mp 1/2$ respectively)
\begin{equation}
\rho^{N,S_L} =
\left( \begin{array}{ll} 0 & 0  \\  0 & 1 \end{array} \right)
\quad\quad
\rho^{N,-S_L} =
\left( \begin{array}{ll} 1 & 0  \\  0 & 0 \end{array} \right) \,,
\label{rhosl}
\end{equation}
so that from Eq. (\ref{rhoq}) we have the quark helicity density matrix
\begin{equation}
\rho^{q/N,S_L} = \left(
\begin{array}{ll} P_A^{q/N,S_L} &   \quad 0  \\
                    \quad 0     & P_P^{q/N,S_L} \end{array} \right) \,.
\label{rhoqsl}
\end{equation}
By using Eqs. (\ref{fr}), (\ref{par}) and (\ref{rhoqsl}) into
Eqs. (\ref{v11})-(\ref{v-10}) we obtain the non zero matrix elements:
\begin{eqnarray}
\rho_{1,1}^{(S_L)}(V) \, d^3\sigma &=&
\sum_q \int \frac {dx}{\pi z} \, f_{q/N} \, d\hat\sigma^q
\left[ P_A^{q/N,S_L} D_{V_1/q_+}
     + P_P^{q/N,S_L} D_{V_1/q_-} \right] \label{vl11} \\
\rho_{0,0}^{(S_L)}(V) \, d^3\sigma &=&
\sum_q \int \frac {dx}{\pi z} \, f_{q/N} \, d\hat\sigma^q \,
D_{V_0/q_+} \label{vl00} \\
\rho_{-1,-1}^{(S_L)}(V) \, d^3\sigma &=&
\sum_q \int \frac {dx}{\pi z} \, f_{q/N} \, d\hat\sigma^q
\left[ P_A^{q/N,S_L} D_{V_1/q_-}
     + P_P^{q/N,S_L} D_{V_1/q_+} \right] \label{vl-1-1}
\end{eqnarray}
where the apex $(S_L)$ reminds of the nucleon spin configuration.

\vskip 6pt
\noindent
{\it b) Nucleon transverse polarization}, $S = S_S$

In this case we have
\begin{equation}
\rho^{N,S_S} =
{1 \over 2} \left( \begin{array}{ll} 1 & 1  \\  1 & 1 \end{array} \right)
\quad\quad
\rho^{N,-S_S} =
{1 \over 2} \left( \begin{array}{ll} \phantom {-} 1 & -1 \\
                   -1 & \phantom{-} 1 \end{array} \right)
\label{rhoss}
\end{equation}
which, via Eqs. (\ref{rhoq}) and (\ref{polq}), imply
\begin{equation}
\rho^{q/N,S_S} = {1 \over 2} \left(
\begin{array}{ll} \quad 1 & P^{q/N,S_S} \\ P^{q/N,S_S} & \quad 1 \end{array}
\right) \,.
\label{rhoqss}
\end{equation}

Insertion of Eq. (\ref{rhoqss}) into Eqs. (\ref{v11})-(\ref{v-10})
now obtains both diagonal and non diagonal matrix elements; the diagonal
ones are
\begin{eqnarray}
\rho_{1,1}^{(S_T)}(V) \, d^3\sigma &=&
\sum_q \int \frac {dx}{\pi z} \, f_{q/N} \, d\hat\sigma^q \,
{1 \over 2} \left[ D_{V_1/q_+} + D_{V_1/q_-} \right] \label{vs11} \\
\rho_{0,0}^{(S_T)}(V) \, d^3\sigma &=&
\sum_q \int \frac {dx}{\pi z} \, f_{q/N} \, d\hat\sigma^q \,
D_{V_0/q_+} \label{vs00} \\
\rho_{-1,-1}^{(S_T)}(V) &=& \rho_{1,1}^{(S_T)}(V) \label{vs-1-1}
= {1 - \rho_{0,0}^{(S_T)}(V) \over 2}
\end{eqnarray}
where we have used an apex $S_T$, rather than $S_S$,
because the same results, as we shall immediately see, hold also in the
other transverse spin case, $S=S_N$.

The non zero non diagonal matrix elements are
\begin{eqnarray}
\rho_{1,0}^{(S_S)}(V) \, d^3\sigma &=&
\sum_q \int \frac {dx}{\pi z} \, f_{q/N} \, {P^{q/N,S_S} \over 2}
\left[ \mbox{\rm Re} \hat M_+^q \hat M_-^{q*} \right]
D_{1,0}^{+,-} \label{vs10} \\
\rho_{-1,0}^{(S_S)}(V) &=& \rho_{1,0}^{(S_S)}(V) \label{vs-10}
\end{eqnarray}
which involve the non diagonal fragmentation functions (\ref{framp}).

\vskip 6pt
\noindent
{\it c) Nucleon transverse polarization}, $S =  S_N$

We now have the nucleon helicity density matrices
\begin{equation}
\rho^{N,S_N} =
{1 \over 2} \left( \begin{array}{ll} \phantom{-} 1 & i  \\
-i & 1 \end{array} \right)
\quad\quad
\rho^{N,-S_N} =
{1 \over 2} \left( \begin{array}{ll} 1 & -i \\ i & \phantom{-} 1
\end{array} \right)
\label{rhosn}
\end{equation}
which lead to
\begin{equation}
\rho^{q/N,S_N} = {1 \over 2} \left(
\begin{array}{ll} \quad \> 1 & iP^{q/N,S_N}  \\
               -iP^{q/N,S_N} &   \quad 1 \end{array} \right) \,.
\label{rhoqsn}
\end{equation}

Insertion of Eq. (\ref{rhoqsn}) into Eqs. (\ref{v11})-(\ref{v-10}) gives
the same results as those obtained with $S=S_S$ for the diagonal matrix
elements, Eqs. (\ref{vs11})-(\ref{vs-1-1}). The non zero non diagonal
matrix elements are instead given by
\begin{eqnarray}
\rho_{1,0}^{(S_N)}(V) \, d^3\sigma &=&
\sum_q \int \frac {dx}{\pi z} f_{q/N} \, {iP^{q/N,S_N} \over 2}
\left[ \mbox{\rm Re} \hat M_+^q \hat M_-^{q*} \right]
D_{1,0}^{+,-} \label{vn10} \\
\rho_{-1,0}^{(S_N)}(V) &=& - \rho_{1,0}^{(S_N)}(V) \,. \label{vn-10}
\end{eqnarray}
Notice that, by rotational invariance, $P^{q/N,S_N} = P^{q/N,S_S}$, so that
$\rho_{1,0}^{(S_N)} = i \rho_{1,0}^{(S_S)}$.

\vskip 12pt
\noindent
{\bf 3 - Spin 1/2 final hadron; unpolarized leptons and polarized nucleons}
\vskip 6pt

In case of final spin 1/2 hadrons ($h=B$), with unpolarized leptons and
spin $S$ nucleons, we have from Eqs. (\ref{gen}), (\ref{for}) and (\ref{norm})
and in analogy to Eqs. (\ref{v11})-(\ref{v-10}):
\begin{eqnarray}
\rho_{+,+}^{(S)}(B) \, d^3\sigma &=&
\sum_q \int \frac {dx}{\pi z} \, f_{q/N} \, d\hat\sigma^q
\left[ \rho_{+,+}^{q/N,S} D_{+,+}^{+,+}
+ \rho_{-,-}^{q/N,S} D_{+,+}^{-,-} \right] \label{b++} \\
\rho_{-,-}^{(S)}(B) \, d^3\sigma &=&
\sum_q \int \frac {dx}{\pi z} \, f_{q/N} \, d\hat\sigma^q
\left[ \rho_{+,+}^{q/N,S} D_{-,-}^{+,+}
+ \rho_{-,-}^{q/N,S} D_{-,-}^{-,-} \right] \label{b--} \\
\rho_{+,-}^{(S)}(B) \, d^3\sigma &=&
\sum_q \int \frac {dx}{\pi z} \, f_{q/N}
\left[ \mbox{\rm Re} \hat M_+^q \hat M_-^{q*} \right]
\rho_{+,-}^{q/N,S} D_{+,-}^{+,-} \label{b+-}
\end{eqnarray}
with $d^3\sigma$ given by Eq. (\ref{sunp}).
Notice that $\rho_{+,+} + \rho_{-,-} = 1, \> \rho_{-,+} = \rho_{+,-}^*$
and that, from Eqs. (\ref{def}) and (\ref{par}), $D_{+,-}^{+,-}$ is real.

When considering particular nucleon spin configurations we obtain for
$S=S_L$, in analogy to Eqs. (\ref{vl11})-(\ref{vl-1-1}),
\begin{eqnarray}
\rho_{+,+}^{(S_L)}(B) \, d^3\sigma &=&
\sum_q \int \frac {dx}{\pi z} f_{q/N} \, d\hat\sigma^q
\left[ P_A^{q/N,S_L} D_{B_+/q_+}
     + P_P^{q/N,S_L} D_{B_+/q_-} \right] \label{bl++} \\
\rho_{-,-}^{(S_L)}(B) \, d^3\sigma &=&
\sum_q \int \frac {dx}{\pi z} f_{q/N} \, d\hat\sigma^q
\left[ P_A^{q/N,S_L} D_{B_+/q_-}
     + P_P^{q/N,S_L} D_{B_+/q_+} \right] \label{bl--}
\end{eqnarray}
whereas, in the transverse spin cases, we have for the diagonal
matrix elements
\begin{equation}
\rho_{+,+}^{(S_T)}(B) = \rho_{-,-}^{(S_T)}(B) = {1\over 2}
\label{btd}
\end{equation}
and for the off-diagonal ones, in analogy to Eqs. (\ref{vs10})
and (\ref{vn10}),
\begin{eqnarray}
\rho_{+,-}^{(S_S)}(B) \, d^3\sigma &=&
\sum_q \int \frac {dx}{\pi z} f_{q/N} \, {P^{q/N,S_S} \over 2}
\left[ \mbox{\rm Re} \hat M_+^q \hat M_-^{q*} \right]
D_{+,-}^{+,-} \label{bs+-} \\
\rho_{+,-}^{(S_N)}(B) \, d^3\sigma &=&
\sum_q \int \frac {dx}{\pi z} f_{q/N} \, {iP^{q/N,S_N} \over 2}
\left[ \mbox{\rm Re} \hat M_+^q \hat M_-^{q*} \right]
D_{+,-}^{+,-} \,. \label{bn+-}
\end{eqnarray}

The knowledge of the helicity density matrix $\rho(B)$ allows to compute the
expectation values of the components of the polarization vector $\bfP(B)$,
in the {\it helicity rest frame} of $B$ \cite{bou}:
\begin{equation}
P_i = {\mbox{\rm Tr}}(\sigma^i \rho) \,,
\label{tr}
\end{equation}
which yields
\begin{eqnarray}
P_x^{(S_S)} \, d^3\sigma &=&
\sum_q \int \frac {dx}{\pi z} \, f_{q/N} \, P^{q/N,S_S}
\left[ \mbox{\rm Re} \hat M_+^q \hat M_-^{q*} \right]
D_{+,-}^{+,-} \label{px} \\
P_y^{(S_N)} &=& - P_x^{(S_S)} \label{py} \\
P_z^{(S_L)} \, d^3\sigma &=&
\sum_q \int \frac {dx}{\pi z} \, f_{q/N} \, P^{q/N,S_L} \,
d\hat\sigma^q \, \left[ D_{B_+/q_-} - D_{B_+/q_+} \right] \label{pz}
\end{eqnarray}
where we have used Eq. (\ref{polq}) and the fact that, by parity invariance,
$D_{+,-}^{+,-}$ is real. All other components of the polarization vectors are
zero. Let us stress once more that the $x,y$ and $z$ components in the above
equations (\ref{px})-(\ref{pz}) refer to the coordinate axes in the helicity
rest frame of hadron $B$, whereas the apices $S_L, S_N$ and $S_S$ are related
to the nucleon spin orientations in the reference frame where we compute
the scattering [see comments after Eq. (\ref{sunp})].

The quantities in the right hand sides of Eqs. (\ref{px})-(\ref{pz}) might
look more familiar if written in different notations or in different spin
basis. In fact we have
\begin{equation}
f_{q/N}(x) \, P_{P(A)}^{q/N,S_L}(x) = f_{q_{+(-)}/N_+}(x) = f_{q_{-(+)}/N_-}(x)
\label{pold}
\end{equation}
where $f_{q_{+(-)}/N_+}$ is the polarized distribution function, that is
the density number of quarks with helicity $+(-)$ inside a nucleon with
helicity + and the last equality holds due to parity invariance.
{}From Eq. (\ref{polq}) and Eq. (\ref{pold}) one has
\begin{equation}
f_{q/N} \, P^{q/N,S_L} = f_{q_+/N_+} - f_{q_-/N_+} \equiv \Delta q
\label{dq}
\end{equation}
and similarly for spin quantized along a transverse direction $T=N,S$,
\begin{equation}
f_{q/N} \, P^{q/N,S_T} = f_{q,S_T/N,S_T} - f_{q,-S_T/N,S_T}
\equiv \Delta_T q \,.
\label{dtq}
\end{equation}
By switching from the helicity to the $N$ spin quantization basis (notice
that the $N$ direction, {\it i.e.} the $y$-axis, is the same both for the
initial nucleon and the final hadron helicity rest frame) one obtains
[see Eq. (\ref{not})]:
\begin{equation}
- \left[ \mbox{\rm Re} \hat M_+^q \hat M_-^{q*} \right]
= {d\hat\sigma^{\ell + q,S_N \to \ell + q,S_N} \over d\hat t}
- {d\hat\sigma^{\ell + q,S_N \to \ell + q,-S_N}\over d\hat t}
\equiv \Delta_N \hat\sigma^q
\label{nbas}
\end{equation}
and
\begin{equation}
D_{+,-}^{+,-} = D_{B,S_N/q,S_N} - D_{B,-S_N/q,S_N}
\equiv \Delta_T D_{B/q} \,,
\label{dtd}
\end{equation}
which is a difference of transverse fragmentation functions and is the same
for any transverse spin direction $T=N,S$. Similarly, one defines
\begin{equation}
D_{B,S_L/q,S_L} - D_{B,-S_L/q,S_L} = D_{B_+/q_+} - D_{B_-/q_+}
\equiv \Delta D_{B/q} \,.
\label{dd}
\end{equation}

Eqs. (\ref{px})-(\ref{pz}) then read
\begin{eqnarray}
P_x^{(S_S)} &=& - P_y^{(S_N)} \label{px'} \\
P_y^{(S_N)} \, d^3\sigma &=&
\sum_q \int \frac {dx}{\pi z} \, \Delta_T q \> \Delta_N \hat\sigma^q \>
\Delta_T D_{B/q} \label{py'} \\
P_z^{(S_L)} \, d^3\sigma &=&
- \sum_q \int \frac {dx}{\pi z} \, \Delta q \> d\hat\sigma^q \>
\Delta D_{B/q} \label{pz'} \,.
\end{eqnarray}

\vskip 12pt
\noindent
{\bf 4 - Longitudinally polarized leptons and polarized nucleons}
\vskip 6pt

We discuss now the case of polarized leptons; as we noticed after
Eq. (\ref{gen}) only the diagonal elements of the lepton helicity density
matrix $\rho^{\ell,s}$ contribute to $\rho(h)$, so that only longitudinal
polarizations could affect the results. We consider then longitudinally
polarized leptons, $s=s_L$, which amounts to
\begin{equation}
\rho_{\lambda^{\,}_{\ell},\lambda^{\prime}_{\ell}}^{\ell,s_L} =
\delta_{\lambda^{\,}_{\ell},\lambda^{\prime}_{\ell}} \>
\delta_{\lambda^{\,}_{\ell},+} \,.
\label{rlon}
\end{equation}
Eq. (\ref{gen}) then reads
\begin{eqnarray}
\rho_{\lambda^{\,}_h,\lambda^\prime_h}^{(s_L,S)}(h) \>
\frac{E_h \, d^3\sigma^{\ell,s_L + N,S \to h + X}} {d^{3} \bfp_h} &=&
\sum_{q; \lambda^{\,}_q, \lambda^\prime_q}
\int \frac {dx}{\pi z} \frac {1}{16 \pi x^2 s^2} \label{genpl} \\
& & \rho_{\lambda^{\,}_q, \lambda^{\prime}_q}^{q/N,S} \, f_{q/N}(x) \,
\hat M^q_{+, \lambda^{\,}_q; +, \lambda^{\,}_q} \,
\hat M^{q*}_{+, \lambda^{\prime}_q; +, \lambda^{\prime}_q} \,
D_{\lambda^{\,}_h, \lambda^{\prime}_h}^{\lambda^{\,}_q,\lambda^{\prime}_q}(z)
\nonumber
\end{eqnarray}
which holds for any final hadron and any nucleon spin orientation.
We briefly discuss the same cases treated in the previous two Sections;
for convenience we use the notations defined in Eqs. (\ref{pold})-(\ref{dtq}).

\vskip 6pt
\noindent
i) $h=V$, $S=S_L$

Eqs. (\ref{genpl}), (\ref{rhosl}) and (\ref{not}) give [compare with
Eqs. (\ref{vl11})-(\ref{vl-1-1})]
\begin{eqnarray}
\!\!\!\!\!\!
\rho_{1,1}^{(s_L,S_L)}(V) \, d^3\sigma_L &=&
\sum_q \int \frac {dx}{\pi z}
\left[ f_{q_-/N_+} \, |\hat M_+^q|^2 \, D_{V_1/q_+}
     + f_{q_+/N_+} \, |\hat M_-^q|^2 \, D_{V_1/q_-}
\right] \label{vl11pl} \\
\!\!\!\!\!\!\!
\rho_{0,0}^{(s_L,S_L)}(V) \, d^3\sigma_L &=&
\sum_q \int \frac {dx}{\pi z}
\left[ f_{q_-/N_+} \, |\hat M_+^q|^2
     + f_{q_+/N_+} \, |\hat M_-^q|^2 \right] \, D_{V_0/q_+}
\label{vl00pl} \\
\!\!\!\!\!\!\!
\rho_{-1,-1}^{(s_L,S_L)}(V) \, d^3\sigma_L &=&
\sum_q \int \frac {dx}{\pi z}
\left[ f_{q_-/N_+} \, |\hat M_+^q|^2 \, D_{V_1/q_-}
     + f_{q_+/N_+} \, |\hat M_-^q|^2 \, D_{V_1/q_+}
\right] \label{vl-1-1pl}
\end{eqnarray}
where the apex $(s_L,S_L)$ reminds of the lepton and nucleon spin
configurations and $d^3\sigma_L$ stands for
\begin{equation}
\frac{E_V \, d^3\sigma^{\ell,s_L + N,S_L \to V + X}} {d^{3} \bfp_V} =
\sum_q \int \frac {dx}{\pi z}
\left[ f_{q_-/N_+} |\hat M_+^q|^2 + f_{q_+/N_+} |\hat M_-^q|^2 \right]
D_{V/q} \,. \label{normpl}
\end{equation}

\vskip 6pt
\noindent
ii) $h=V$, $S=S_T \>(T=S,N)$

Eqs. (\ref{vs11})-(\ref{vs-1-1}) now modify into
\begin{eqnarray}
\rho_{1,1}^{(s_L,S_T)}(V) \, d^3\sigma &=&
\sum_q \int \frac {dx}{\pi z} \, f_{q/N} \, {1 \over 2}
\left[ |\hat M_+^q|^2 \, D_{V_1/q_+} + |\hat M_-^q|^2 \, D_{V_1/q_-} \right]
\label{vs11pl} \\
\rho_{0,0}^{(s_L,S_T)}(V) \, d^3\sigma &=&
\sum_q \int \frac {dx}{\pi z} \, f_{q/N} \, d\hat\sigma^q \,
D_{V_0/q_+} \label{vs00pl} \\
\rho_{-1,-1}^{(s_L,S_T)}(V) \, d^3\sigma &=&
\sum_q \int \frac {dx}{\pi z} \, f_{q/N} \, {1 \over 2}
\left[ |\hat M_+^q|^2 \, D_{V_1/q_-} + |\hat M_-^q|^2 \, D_{V_1/q_+} \right]
\label{vs-1-1pl}
\end{eqnarray}
where $d^3\sigma$ is the unpolarized cross-section (\ref{sunp}),
and the non diagonal matrix elements are
\begin{eqnarray}
\rho_{1,0}^{(s_L,S_S)}(V) \, d^3\sigma &=&
\sum_q \int \frac {dx}{\pi z} \, {\Delta_T q \over 2} \>
\hat M_+^q \hat M_-^{q*} \> D_{1,0}^{+,-} \label{vs10pl} \\
\rho_{-1,0}^{(s_L,S_S)}(V) \, d^3\sigma &=&
\sum_q \int \frac {dx}{\pi z} \, {\Delta_T q \over 2} \>
\hat M_-^q \hat M_+^{q*} \> D_{1,0}^{+,-} \label{vs-10pl} \\
\rho_{1,0}^{(s_L,S_N)}(V) \, d^3\sigma &=&
\sum_q \int \frac {dx}{\pi z} \, {i\Delta_T q \over 2} \>
\hat M_+^q \hat M_-^{q*} \> D_{1,0}^{+,-} \label{vn10pl} \\
\rho_{-1,0}^{(s_L,S_N)}(V) \, d^3\sigma &=&
- \sum_q \int \frac {dx}{\pi z} \, {i\Delta_T q \over 2} \>
\hat M_-^q \hat M_+^{q*} \> D_{1,0}^{+,-} \,. \label{vn-10pl}
\end{eqnarray}
The non diagonal elements $\rho_{\pm1,0}$ might differ from
those found with unpolarized leptons, Eqs. (\ref{vs10}), (\ref{vs-10}),
(\ref{vn10}) and (\ref{vn-10}), only if the amplitude product
$\hat M_+^q \hat M_-^{q*}$ is a complex quantity, which is certainly
not the case at lowest perturbative order. Notice also that
$\rho_{\pm 1,0}^{(s_L,S_N)} = \pm i \rho_{\pm 1,0}^{(s_L,S_S)}$.

\vskip 6pt
\noindent
iii) $h=B$, $S=S_L$

Longitudinally polarized leptons and longitudinally polarized nucleons
lead to final spin half hadrons with
\begin{eqnarray}
\!\!\!\!\!\!\!\!
\rho_{+,+}^{(s_L,S_L)}(B) \, d^3\sigma_L &=&
\sum_q \int \frac {dx}{\pi z}
\left[ f_{q_-/N_+} \, |\hat M_+^q|^2 \, D_{B_+/q_+}
     + f_{q_+/N_+} \, |\hat M_-^q|^2 \, D_{B_+/q_-}
\right] \label{bl++pl} \\
\!\!\!\!\!\!\!\!
\rho_{-,-}^{(s_L,S_L)}(B) \, d^3\sigma_L &=&
\sum_q \int \frac {dx}{\pi z}
\left[ f_{q_-/N_+} \, |\hat M_+^q|^2 \, D_{B_+/q_-}
     + f_{q_+/N_+} \, |\hat M_-^q|^2 \, D_{B_+/q_+}
\right] \label{bl--pl}
\end{eqnarray}
whereas with transversely polarized nucleons one has

\vskip 6pt
\noindent
iv) $h=B$, $S=S_T$

\begin{eqnarray}
\!\!\!\rho_{+,+}^{(s_L,S_T)}(B) \, d^3\sigma &=&
\sum_q \int \frac {dx}{\pi z} \, f_{q/N} \, {1 \over 2}
\left[ |\hat M_+^q|^2 \, D_{B_+/q_+} + |\hat M_-^q|^2 \, D_{B_+/q_-}
\right] \label{bt++pl} \\
\!\!\!\rho_{-,-}^{(s_L,S_T)}(B) \, d^3\sigma &=&
\sum_q \int \frac {dx}{\pi z} \, f_{q/N} \, {1 \over 2}
\left[ |\hat M_+^q|^2 \, D_{B_+/q_-} + |\hat M_-^q|^2 \, D_{B_+/q_+}
\right] \label{bt--pl} \\
\!\!\!\rho_{+,-}^{(s_L,S_S)}(B) \, d^3\sigma &=&
\sum_q \int \frac {dx}{\pi z} \> {\Delta_T q \over 2} \>
\hat M_+^q \hat M_-^{q*} \> D_{+,-}^{+,-} \label{bs+-pl} \\
\!\!\!\rho_{+,-}^{(s_L,S_N)}(B) \, d^3\sigma &=&
\sum_q \int \frac {dx}{\pi z} \> {i\Delta_T q \over 2} \>
\hat M_+^q \hat M_-^{q*} \> D_{+,-}^{+,-} \,. \label{bn+-pl}
\end{eqnarray}
Again, in comparison to the case of unpolarized leptons
[see Eqs. (\ref{bl++})-(\ref{bn+-})], one finds different results for
the diagonal matrix elements, but the same ones, provided
$\hat M_+^q \hat M_-^{q*}$ is real, for the non diagonal elements.
Similarly to Eqs. (\ref{px})-(\ref{pz})
or (\ref{px'})-(\ref{pz'}), the above equations can also be written in
terms of components of the polarization vector of hadron $B$:
\begin{eqnarray}
P_x^{(s_L,S_S)} &=& - P_y^{(s_L,S_N)} =
P_x^{(S_S)} = -P_y^{(S_N)} \label {pxpl} \\
P_z^{(s_L,S_L)} \, d^3\sigma &=&
\sum_q \int \frac {dx}{\pi z}
\left[ f_{q_-/N_+} \, |\hat M_+^q|^2 - f_{q_+/N_+} \, |\hat M_-^q|^2 \right]
\Delta D_{B/q} \label{pzlpl} \\
P_z^{(s_L,S_T)} \, d^3\sigma &=&
\sum_q \int \frac {dx}{\pi z} \, f_{q/N} \, {1 \over 2}
\left[ |\hat M_+^q|^2 -  |\hat M_-^q|^2 \right] \Delta D_{B/q} \,.
\label{pztpl}
\end{eqnarray}

\vskip 12pt
\noindent
{\bf 5 - Possible measurements}
\vskip 6pt

After the theoretical analysis of the previous Sections we should now
address the question of which elements of the helicity density matrix of the
produced hadrons can be measured; we discuss in details the cases of
spin 1 $\rho$ vector mesons and spin 1/2 $\Lambda$ baryons as the most
typical and simple ones, but the same procedure could be applied to other
hadrons.

$\rho$ particles decay into two pions and, in order to measure the helicity
density matrix of the decaying $\rho$, one has to look at the angular
distribution of either one of the produced pions; such a decay is
given by \cite{bou}
\begin{eqnarray}
W(\theta_{\pi},\phi_{\pi}) &=& {3\over 4\pi} \Biggl \{ {1 \over 2}
(1 - \rho_{0,0})
+ {1 \over 2} (3\rho_{0,0} - 1) \cos^2\theta_{\pi} \nonumber \\
&-& {1 \over \sqrt 2} \sin 2\theta_{\pi} \cos\phi_{\pi} \,
{\mbox{\rm Re}}[\rho_{1,0} - \rho_{-1,0}^*]
+ {1 \over \sqrt 2} \sin 2\theta_{\pi} \sin\phi_{\pi} \,
{\mbox{\rm Im}}[\rho_{1,0} - \rho_{-1,0}^*] \nonumber \\
&-& \sin^2\theta_{\pi}\cos2\phi_{\pi} \, {\mbox{\rm Re}}[\rho_{1,-1}]
+ \sin^2\theta_{\pi}\sin2\phi_{\pi} \, {\mbox{\rm Im}}[\rho_{1,-1}] \Biggr \}
\label{decv}
\end{eqnarray}
where $\theta_{\pi}$ and $\phi_{\pi}$ are respectively the polar and azimuthal
angles of the pion in the helicity rest frame of the $\rho$.

{}From our previous results we see that the above distribution simplifies, in
that several matrix elements are zero, depending on the different spin
configurations of the initial particles. Let us consider the case of
unpolarized leptons with different nucleon spin orientations; we then expect
angular distributions of the following types [see Eqs.
(\ref{vl11})-(\ref{vl-1-1}), (\ref{vs11})-(\ref{vs-10}) and
(\ref{vn10})-(\ref{vn-10})]:

\vskip 6pt
\noindent
{\it a) Nucleon longitudinal polarization}, $S = S_L$
\begin{equation}
W(\theta_{\pi},\phi_{\pi}) = {3\over 4\pi} \left \{ {1 \over 2}
(1 - \rho^{(S_L)}_{0,0})
+ {1 \over 2} (3\rho^{(S_L)}_{0,0} - 1) \cos^2\theta_{\pi}  \right \} \,;
\label{decvl}
\end{equation}

\vskip 6pt
\noindent
{\it b) Nucleon transverse polarization}, $S = S_S$
\begin{eqnarray}
W(\theta_{\pi},\phi_{\pi}) &=& {3\over 4\pi} \Biggl \{ {1 \over 2}
(1 - \rho^{(S_T)}_{0,0})
+ {1 \over 2} (3\rho^{(S_T)}_{0,0} - 1) \cos^2\theta_{\pi} \nonumber \\
&+& \sqrt 2 \sin 2\theta_{\pi} \sin\phi_{\pi} \,
{\mbox{\rm Im}}\rho^{(S_S)}_{1,0} \Biggr \} \,;
\label{decvs}
\end{eqnarray}

\vskip 6pt
\noindent
{\it c) Nucleon transverse polarization}, $S = S_N$
\begin{eqnarray}
W(\theta_{\pi},\phi_{\pi}) &=& {3\over 4\pi} \Biggl \{ {1 \over 2}
(1 - \rho^{(S_T)}_{0,0})
+ {1 \over 2} (3\rho^{(S_T)}_{0,0} - 1) \cos^2\theta_{\pi} \nonumber \\
&-& \sqrt 2 \sin 2\theta_{\pi} \cos\phi_{\pi} \,
{\mbox{\rm Re}}\rho^{(S_N)}_{1,0} \Biggr \} \,.
\label{decvn}
\end{eqnarray}
Notice that $\rho_{0,0}^{(S_L)} =  \rho_{0,0}^{(S_T)}$ [Eqs. (\ref{vl00}),
(\ref{vs00})] and that, from $\rho_{1,0}^{(S_N)} = i \rho_{1,0}^{(S_S)}$ one
has Re\,$\rho_{1,0}^{(S_N)} = -$ Im\,$\rho_{1,0}^{(S_S)}$; the observation
of the angular distribution of the $\rho \to \pi\pi$ decay supplies then
only information on $\rho^{(S_L)}_{0,0}(V)$ and Re\,$\rho^{(S_N)}_{1,0}$, that
is on the quantities:
\begin{eqnarray}
\rho_{0,0}^{(S_L)}(V) \, d^3\sigma &=&
\sum_q \int \frac {dx}{\pi z} \, f_{q/N} \, d\hat\sigma^q \,
D_{V_0/q_+} \label{vl00'} \\
{\mbox{\rm Re}} \rho_{1,0}^{(S_N)}(V) \, d^3\sigma &=&
- \sum_q \int \frac {dx}{\pi z} \, f_{q/N} \, {P^{q/N,S_T}
\over 2} \left[ \mbox{\rm Re} \hat M_+^q \hat M_-^{q*} \right]
\mbox{\rm Im} D_{1,0}^{+,-} \label{revn10}
\end{eqnarray}
with $d^3\sigma$ given in Eq. (\ref{sunp}).

Similar results hold in case of longitudinally polarized initial leptons,
with the only difference that in such case $\rho_{0,0}^{(s_L,S_L)}$ differs
from $\rho_{0,0}^{(s_L,S_T)}$ [Eqs. (\ref{vl00pl}), (\ref{vs00pl})] and the
two separate measurements might offer more information.

Let us now consider the production of a spin half baryon which, via its
parity violating decay, allows a measurement of its polarization vector;
the most typical example is the $\Lambda \to p \pi^-$ decay. The angular
distribution of the proton as it emerges in the $\Lambda$ helicity rest
frame is given by
\begin{eqnarray}
W(\theta_p,\phi_p) &=&
{1 \over 4\pi} [ 1 + \alpha \cos\theta_p (2 \rho_{+,+} - 1)
+ 2 \alpha \sin\theta_p\cos\phi_p \, \mbox{\rm Re} \rho_{+,-} \nonumber \\
&-& 2 \alpha \sin\theta_p\sin\phi_p \, \mbox{\rm Im} \rho_{+,-}]
\nonumber \\
&=& {1 \over 4\pi} \left[ 1 + \alpha (P_z \cos\theta_p
+ P_x \sin\theta_p\cos\phi_p + P_y \sin\theta_p\sin\phi_p) \right] \nonumber \\
&=& {1 \over 4\pi} \left[ 1 + \alpha \bfP \cdot \hat{\bfp} \right] \label{decb}
\end{eqnarray}
where $\bfP$ is the $\Lambda$ polarization vector and $\hat{\bfp}$ is the
unit vector along the proton direction in the $\Lambda$ helicity rest frame.
The decay parameter $\alpha$ is experimentally known and for the
$\Lambda \to p \pi^-$ decay $\alpha = 0.642 \pm 0.013$.

{}From the results of Eqs. (\ref{px'})-(\ref{pz'}) we expect then, for
unpolarized leptons and longitudinally polarized nucleon $(S=S_L)$,
the $\Lambda$ decay angular distribution
\begin{equation}
W(\theta_p,\phi_p) =
{1 \over 4\pi} \left[ 1 + \alpha \, P^{(S_L)}_z \cos\theta_p  \right] \,;
\label{decbl}
\end{equation}
for unpolarized leptons and $S=S_S$ we have
\begin{equation}
W(\theta_p,\phi_p) =
{1 \over 4\pi} \left[ 1 + \alpha \, P^{(S_S)}_x \sin\theta_p\cos\phi_p
\right] \,; \label{decbs}
\end{equation}
and for $S=S_N$
\begin{equation}
W(\theta_p,\phi_p) =
{1 \over 4\pi} \left[ 1 + \alpha \, P^{(S_N)}_y \sin\theta_p\sin\phi_p
\right] \,. \label{decbn}
\end{equation}
Recalling that $P_x^{(S_S)} = - P_y^{(S_N)}$ a measurement of
$W(\theta_p,\phi_p)$ supplies information on the two quantities given
in Eqs. (\ref{py'}), (\ref{pz'}) or (\ref{px}), (\ref{pz}).

Similar results are obtained when performing experiments with longitudinally
polarized leptons, with the difference that one has a non zero $P_z$-component
also for transversely polarized nucleons; one can then get further information
on the polarized distribution and fragmentation functions via
Eqs. (\ref{pzlpl}) and (\ref{pztpl}).

\goodbreak
\vskip 12pt
\noindent
{\bf 6 - Some numerical estimates and conclusions}
\vskip 6pt
\nobreak

Which values could we expect for the measurable helicity density matrix
elements? Let us consider first spin 1 vector mesons -- $\rho$ particles --
produced with unpolarized leptons and the matrix elements
$\rho^{(S_L)}_{0,0}$ and Re$\rho^{(S_N)}_{1,0}$, Eqs. (\ref{vl00'}) and
(\ref{revn10}) respectively; these are the only independent matrix elements
which can be measured through the decay angular distributions (\ref{decvl})
and (\ref{decvn}).

It is not difficult to give an estimate of $\rho^{(S_L)}_{0,0}(V)$ if one
observes final mesons with large $|x_F|$ and $Q^2$ values, so that one
can safely argue that they contain the original fragmenting quark as a
valence one; if one assumes that, for valence quarks,
$D_{V_0/q_+} = C \, D_{V/q}$, with $C$ a flavour and $z$-independent constant,
then from Eqs. (\ref{vl00'}) and (\ref{sunp}) one obtains
$\rho^{(S_L)}_{0,0}(V) = C$. For $\rho$ particles
and according to $SU(6)$ wave functions, one has $C=1/3$, so that we expect
\begin{equation}
\rho^{(S_L)}_{0,0}(\rho) = {1 \over 3}
\label{r00}
\end{equation}
for $\rho$ mesons produced in a kinematical region dominated by valence
quark hadro\-ni\-zation. The same result holds in case of a production
initiated
by longitudinally polarized leptons, Eqs. (\ref{vl00pl}) and (\ref{normpl}).
Notice that the above value of 1/3 leads to a constant angular decay
distribution $W(\theta_\pi, \phi_\pi) = 1/4\pi$, Eq. (\ref{decvl}), which
should be easily detected experimentally.

The evaluation of Re$\rho^{(S_N)}_{1,0}$ via Eq. (\ref{revn10}) requires
the knowledge of non perturbative fragmentation properties, contained in
Im$D^{+,-}_{1,0}$; actually, we expect a measurement of Re$\rho^{(S_N)}_{1,0}$
to give us information on such a quantity. However, we try here to obtain
an idea of the possible maximum value of Re$\rho^{(S_N)}_{1,0}$, by assuming,
as naively suggested by Eqs. (\ref{framp}) and (\ref{fr}),
\begin{equation}
\mbox{\rm Im}D_{1,0}^{+,-} \simeq \left[ D_{V_1/q_+} \, D_{V_0/q_-}
\right]^{1/2} \,. \label{imv}
\end{equation}
If, again, we consider only $SU(6)$ valence quark contributions to $\rho$
production, we have $D_{\rho^{\,}_0/q_+} = (1/3) \, D_{\rho/q}$ and
$D_{\rho^{\,}_1/q_+} = (2/3) \, D_{\rho/q}$, so that we take
\begin{equation}
\mbox{\rm Im}D_{1,0}^{+,-} \simeq  {\sqrt 2 \over 3} D_{\rho/q} \,. \label{imr}
\end{equation}

In order to evaluate Re$\rho^{(S_N)}_{1,0}$ we also need to know the
value of the transverse quark polarization inside the transversely
polarized nucleon, $P^{q/N,S_T}$. In general this quantity depends
on $x$; if, however, we consider large $p_T$ final mesons originated
by large $x$ proton valence quarks which fragment into large $z$ hadrons
we can assume, according to $SU(6)$ proton wave functions:
\begin{equation}
P^{u_V/p,S_T} = {2 \over 3} \quad\quad P^{d_V/p,S_T} = -{1\over 3}
\label{pvalp}
\end{equation}
independent of $x$. By interchanging $u$ and $d$ one obtains the analogous
results for neutrons. Sea quarks are assumed not to be polarized.

The elementary interaction $\ell q \to \ell q$, computed at lowest perturbative
order, gives
\begin{equation}
\hat M^q_{++,++} = 8 \pi \alpha \, e_q \, {\hat s \over \hat t}
\quad \quad
\hat M^q_{+-,+-} = 8 \pi \alpha \, e_q \, {\hat u \over \hat t}
\label{m+-}
\end{equation}
where $e_q$ is the quark charge in units of the proton charge;
then [recall Eq. (\ref{not})]
\begin{equation}
\mbox{\rm Re}\hat M_+^q \hat M_-^{q*} =  4 \pi \alpha \, e_q^2 \,
{\hat u \over \hat s \hat t^2} \,\cdot
\label{rem+-}
\end{equation}

By inserting Eqs. (\ref{imr})-(\ref{rem+-}) into
Eqs. (\ref{revn10}) and (\ref{sunp}), we obtain, for the process
$\ell + p,S_N \to \rho^+ + X$ and within the above simplifying assumptions:
\begin{equation}
\mbox{\rm Re}\rho^{(S_N)}_{1,0}(\rho^+) = {2 \sqrt2 \, t \over 9} \frac
{\int dx \, x^{-3} f_{u_V/p}(x) \, D_{\rho^+/u}(z)}
{\int dx \, x^{-3} [(t+xu)^2 + t^2] (t + xu)^{-1}
[f_{u/p}(x) + {1 \over 4} f_{\bar d/p}(x)] D_{\rho^+/u}(z)}
\label{re10r+}
\end{equation}
where we have taken $D_{\rho^+/\bar d} = D_{\rho^+/u}$ and we have used
Eqs. (\ref{zet}) and (\ref{man}). Similar expressions can be derived for
the production of $\rho^0$ and $\rho^-$ mesons.

The above result, Eq. (\ref{re10r+}), can only be considered
as an upper estimate of the magnitude of Re$\rho^{(S_N)}_{1,0}(\rho^+)$;
its sign is arbitrary, due to the phase uncertainty in Eq. (\ref{imv}).
Notice, however, that, in case of production of a $\rho^-$, we
find an opposite sign, due to a leading contribution from $d$ quarks and
Eq. (\ref{pvalp}).

In Figures 1-4 we present numerical results obtained from Eq. (\ref{re10r+})
and the analogous ones for $\rho^-$ and $\rho^+$. We plot the values of
Re$\rho^{(S_N)}_{1,0}(\rho^+)$, Re$\rho^{(S_N)}_{1,0}(\rho^0)$ and
Re$\rho^{(S_N)}_{1,0}(\rho^-)$ for different choices
of $\sqrt s$: in Figs. 1 and 2 we show results at $\sqrt s = 23$ GeV,
respectively at fixed $x_F$ as functions of $p_T$ and at fixed $p_T$ as
functions of $x_F$; analogous results are shown in Figs. 3 and 4 at
$\sqrt s = 314$ GeV. $x_F$ is the $x$-Feynman variable in the $\ell-p$ c.m.
frame and $p_T$ is the $\rho$ transverse momentum. We have used the $u$ and
$\bar d$ distribution functions, including their $Q^2$ evolution, given
in Ref. \cite{dist}; for the quark fragmentation functions we have taken
$D_{\rho^+/u}(z) \sim z^{-1}(1-z)^{1.2}$ from a fit of the experimental
data on $D_{\rho^0/u}(z)$ \cite{dat}. We have explicitely checked that
choices of other available distribution and/or fragmentation functions
leave the numerical results essentially unchanged. Figures 1-4 show that
in all cases sizeable and hopefully detectable values of
$|$Re$\rho^{(S_N)}_{1,0}(\rho)|$ are found; we notice once more that only
the magnitudes and the relative signs of the results for $\rho^+$, $\rho^0$
and $\rho^-$ are meaningful.

Let us finally consider the production of a spin 1/2 baryon, say a $\Lambda$
particle. The decay angular distributions (\ref{decbl}) and (\ref{decbn})
allow a measurement of $P_y^{(S_N)}$ and $P_z^{(S_L)}$, given respectively
in Eqs. (\ref{px}), (\ref{py}) and (\ref{pz}) [or (\ref{py'}) and (\ref{pz'})].
A much simplified version of Eq. (\ref{pz'}) has been recently derived
in Ref. \cite{cin}.

According to $SU(6)$ wave function the entire $\Lambda$ polarization is
due to the strange quark, so that the difference of polarized fragmentation
functions in Eq. (\ref{pz}) is different from zero only for $s$ quarks,
$D_{\Lambda_+/s_-} - D_{\Lambda_+/s_+} = - D_{\Lambda/s}$. Then
Eq. (\ref{pz'}) reads
\begin{equation}
P_z^{(S_L)} = -\frac{\int dx \, (\pi z)^{-1} \, \Delta s \,
d\hat\sigma^s \, D_{\Lambda/s}}
{\sum_q \int dx \, (\pi z)^{-1} \, f_{q/N} \, d\hat\sigma^q \, D_{\Lambda/q}}
\label{pzla}
\end{equation}
Such a quantity is expected to be rather small; however, any non zero value
would offer valuable information on the much debated issue of longitudinal
strange quark polarization, $\Delta s$, inside a longitudinally polarized
nucleon \cite{cin}. A similar information on the transverse polarization can
be obtained from a measurement of $P_y^{(S_N)}$ and Eq. (\ref{py'}).

In conclusion, we have shown how a careful analysis of the spin of hadrons
inclusively produced in the DIS scattering of leptons, either polarized or
not, on polarized nucleons might yield further information on the quark
distribution and on the quark fragmentation properties; we have performed
our analysis in the framework of perturbative QCD and the factorization
theorem, giving comprehensive and detailed expressions for measurable
quantities in several different cases. Any measurement of spin observables
would help in understanding subtle non perturbative spin properties
of hadrons, which would otherwise be inaccessible; once enough information
has been gathered, like in the unpolarized case, predictions for other
processes or observables can reliably be made.

\goodbreak
\vskip 12pt
\noindent
{\bf Acknowledgements}
\vskip 6pt
\nobreak

This work has been supported by the European Community under contract
CHRX-CT94-0450; one of us (J.H.) would like to thank the Department of
Theoretical Physics of Torino University for the kind hospitality.


\newpage
\noindent
{\bf Figure captions}
\vskip 12pt
\noindent
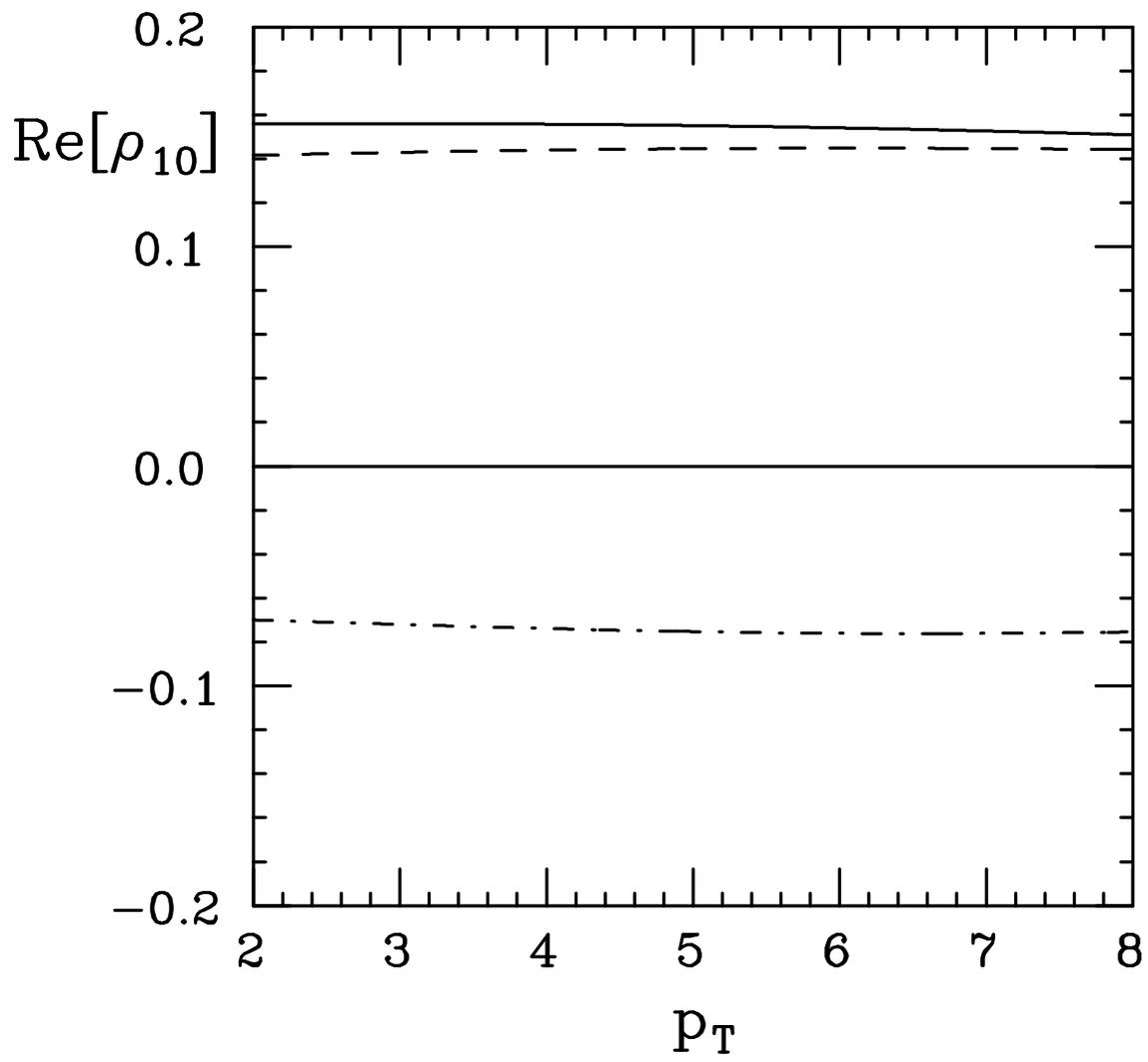
{\bf Fig. 1} -
Plot of Re$\rho^{(S_N)}_{1,0}(\rho^+)$ (solid line),
Re$\rho^{(S_N)}_{1,0}(\rho^0)$ (dashed line) and
Re$\rho^{(S_N)}_{1,0}(\rho^-)$ (dot-dashed line) as functions of $p_T$
at $\sqrt s = 23$ GeV and $x_F = -0.3$. The results are obtained from
Eq. (\ref{re10r+}) of the text and similar equations for $\rho^0$ and
$\rho^-$; the minimum values of $x$ and $Q^2$ contributing are respectively
$x_{min} \simeq 0.33$ and $Q^2_{min} \simeq 4$ (GeV/$c)^2$.
\vskip 6pt
\noindent
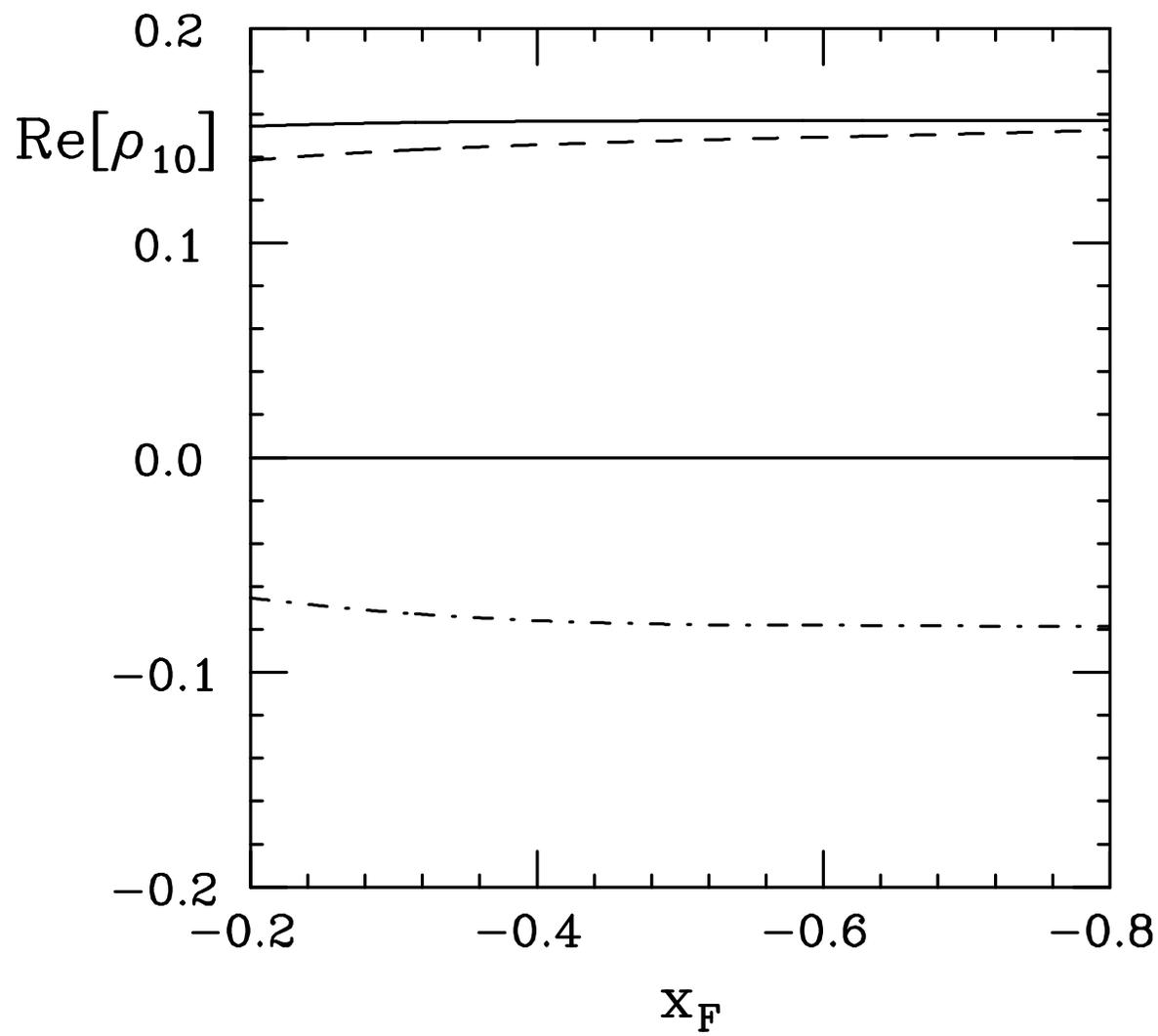
{\bf Fig. 2} -
Plot of Re$\rho^{(S_N)}_{1,0}(\rho^+)$ (solid line),
Re$\rho^{(S_N)}_{1,0}(\rho^0)$ (dashed line) and
Re$\rho^{(S_N)}_{1,0}(\rho^-)$ (dot-dashed line) as functions of $x_F$
at $\sqrt s = 23$ GeV and $p_T = 3$ GeV/$c$. The results are obtained from
Eq. (\ref{re10r+}) of the text and similar equations for $\rho^0$ and
$\rho^-$; the minimum values of $x$ and $Q^2$ contributing are respectively
$x_{min} \simeq 0.28$ and $Q^2_{min} \simeq 10$ (GeV/$c)^2$.
\vskip 6pt
\noindent
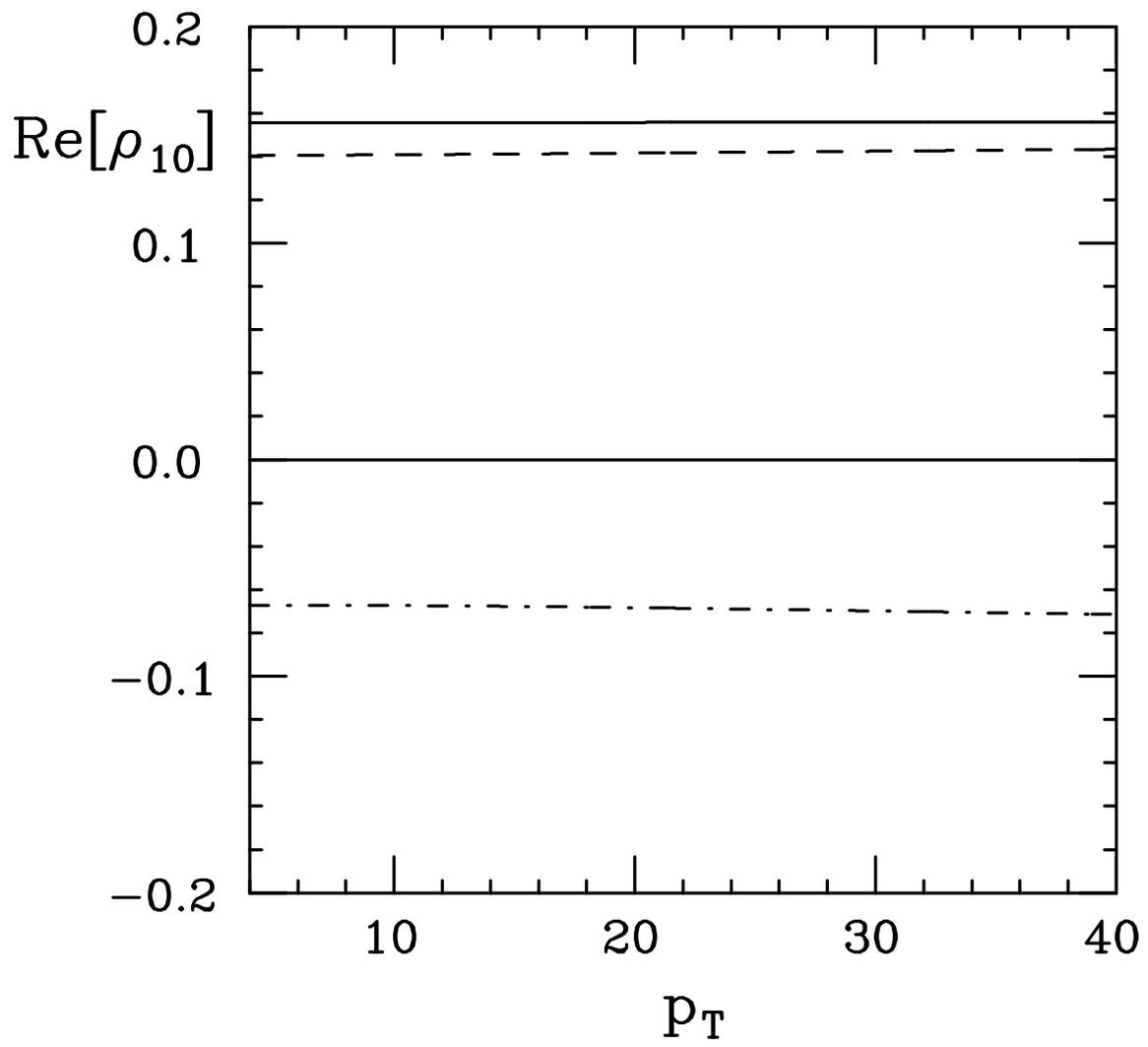
{\bf Fig. 3} -
Same as Fig. 1 at $\sqrt s = 314$ GeV and $x_F = -0.3$;
$x_{min} \simeq 0.30$ and $Q^2_{min} \simeq 16$ (GeV/$c)^2$.
\vskip 6pt
\noindent
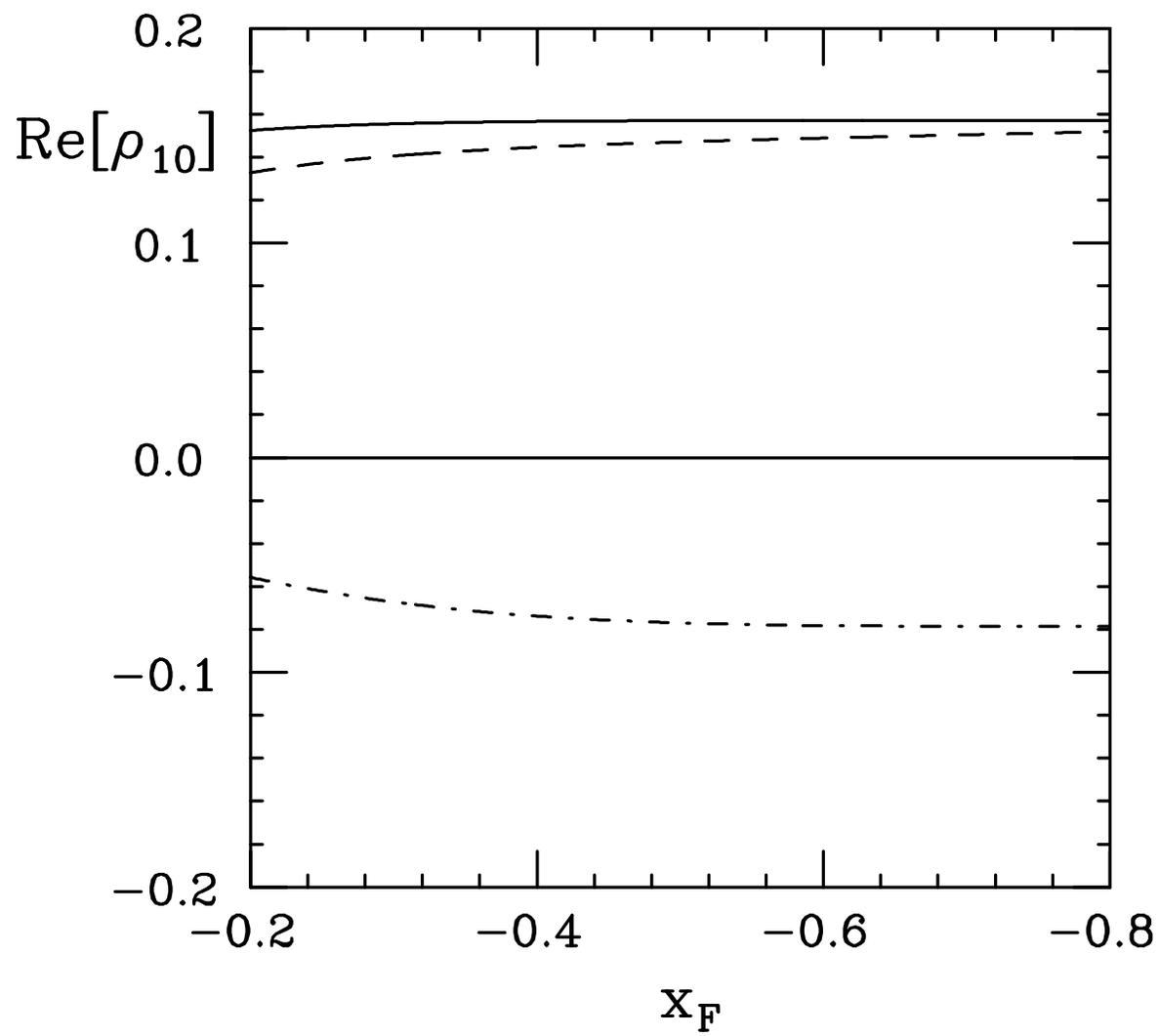
{\bf Fig. 4} -
Same as Fig. 2 at $\sqrt s = 314$ GeV and $p_T = 6$ GeV/$c$;
$x_{min} \simeq 0.20$ and $Q^2_{min} \simeq 36$ (GeV/$c)^2$.
\end{document}